\documentclass{article}    
\usepackage{tikz}
\usepackage{float}
\usepackage[margin=1.4in]{geometry}
\usetikzlibrary{calc}
\usepackage{amssymb}
\usepackage{amsmath}
\usepackage{mathtools}
\usetikzlibrary{automata, positioning, arrows} \newtheorem{theorem}{Theorem} \newtheorem{definition}{Definition}
\usepackage{array}
\usetikzlibrary{decorations.pathreplacing,calc}
\tikzset{%
 middle dotted line/.style={
   decoration={show path construction,
     lineto code={
         \draw[#1] (\tikzinputsegmentfirst) --($(\tikzinputsegmentfirst)!.3333!(\tikzinputsegmentlast)$);,
         \draw[dotted,#1] ($(\tikzinputsegmentfirst)!.3333!(\tikzinputsegmentlast)$)--($(\tikzinputsegmentfirst)!.6666!(\tikzinputsegmentlast)$);,
         \draw[#1] ($(\tikzinputsegmentfirst)!.6666!(\tikzinputsegmentlast)$)--(\tikzinputsegmentlast);,
     }
   },
   decorate
 },
}
\usepackage{multirow}
\usepackage[hyphenbreaks]{breakurl}
\usepackage[hyphens]{url}

\usepackage[utf8]{inputenc}
\usepackage{dsfont}

\usepackage[symbol]{footmisc}

\usepackage[super]{natbib}
\bibpunct{[}{]}{,}{s}{}{;}

    \makeatletter
\def\@fnsymbol#1{\ensuremath{\ifcase#1\or \dagger\or \ddagger\or
   \mathsection\or \mathparagraph\or \|\or **\or \dagger\dagger
   \or \ddagger\ddagger \else\@ctrerr\fi}}
    \makeatother

\begin{document}

\title{A game theoretic approach to lowering incentives to violate speed limits in Finland}
\author{Mika Sutela\thanks{University of Eastern Finland: Law School. Contact by email: \texttt{mika.sutela@uef.fi}} \and Nino Lindström\thanks{University of Helsinki: Institute of Criminology and Legal Policy / Helsinki Graduate School of Economics. Contact by email: \texttt{nino.lindstrom@helsinki.fi}}}

\maketitle

\begin{abstract}
We expand on earlier research on the topic by discussing an infinitely-repeated-game model with a subgame perfect equilibrium strategy profile (SPE) as a solution concept that diminishes incentives to violate speed limits in a \emph{carrot-and-stick} fashion. In attempts to construct an SPE strategy profile, the initial state is chosen such that the drivers are playing a mixed strategy whereas the police is not enforcing with certainty. We also postulate a ``short-period'' version of the repeated game with generalized stage-game payoffs. For this game, we construct a multi-stage strategy profile that is a Nash equilibrium but not an SPE. Some solution candidates are excluded by showing that they do not satisfy a one-shot-deviation property that is a necessary condition for an SPE profile in a repeated game of perfect information.
\end{abstract}

\vspace{1em}

\begin{center}
\textmd{\small INTRODUCTION}
\end{center}

As already Ross (1960) \cite{ross} wrote, traffic law violations are a costly and widespread form of criminal activity, illegal acts which are not stigmatized as criminal by the public. In public opinion, it is not considered "real" criminality, and it has generally not been treated as a subject for criminological study. (Ross, 1960) \cite{ross}. It is not obvious why many injurious acts are regarded as crimes when others are not. Some illegal acts, such as many traffic violations, can be victimless (see Dnes, 2000: 70.)\cite{dnes00}. Traffic offences are crimes \emph{mala quia prohibita}, and therefore, are not in themselves morally objectionable like traditional crimes. Every road user is a potential criminal (Tolvanen, 1999: 61.) \cite{tolvanen}. At least in England, "anti-social traffic behaviour" has been used as a target of policing (see Corbett, 2008: 135; \cite{corbett08} Huotari, 2020: 18, 130; \cite{huotari}). Also speeding has been socially constructed as almost a ”non-crime” or as not ”real” crime (Corbett, 2000)\cite{corbett00}. The speeding behaviour is pervasive, and arguably socially acceptable (Fleiter $\&$ Watson, 2005). Speeding is a socio-cultural problem and cannot be located solely with the individual driver (see Corbett et al., 1998;\cite{corbett98}). Unfortunately speeding has not been viewed as an anti-social activity as many drivers do not perceive speeding as having negative consequences or being a serious offence (Webster $\&$ Wells, 2000.)\cite{webster00}. Road traffic is a rich phenomena which can be analysed from many disciplines point of view. It includes e.g. educational, legal, medical, psychological, sociological and technical aspects. Driver’s behaviour are influenced by several factors, such as traffic rules, different costs (the costs of breaking them, the costs of safety measures decided by the motorist himself, as well as the costs of crashes) as well as external safety factors (Tammi, 2013)\cite{tammi13}. Heuristics have an important role in the field of behavioural economics. They influence people’s decision-making in everyday situations. One heuristic, relevant considering traffic environment, is the availability heuristic. It describes the fact that people assess probabilities according to the mental availability of the corresponding events. For example, if severe traffic offences and fatal road traffic crashes are reported in media more frequently, people become familiar with the topic. Some dramatic event will temporarily alter your feelings about the traffic safety. Fatal crashes are able to bring that topic faster and easier to mind, and therefore, it is estimated to be more likely to occur. The availability heuristic also considers that the assessment of the probabilities of threats (crash or sanction for a traffic crime) depends on individual experiences and observations. Persons who have already got a sanction or involved in a crash, estimate a higher likelihood of becoming a victim of crash or getting a sanction than persons who have been spared so far.  The reason for that, according to the availability heuristic, is the ease with which the memories about the crash or crime are mentally available. In contrast, persons who have not been involved in crashes or captured for a traffic crime in the past, only have an abstract idea of being a victim of a crash or a suspect of criminal act in traffic. Personal experiences, pictures, and vivid examples are more available than incidents that happened to others, or mere words, or statistics. The availability heuristic can lead to probability estimation errors. It is typical for these errors that the probabilities of the occurrence of harmful events, such as sanctions for speeding or traffic crashes, are underestimated (see Tversky $\&$ Kahneman, 1982;\cite{tversky82} Kahneman, 2012;\cite{kahneman12} Schulan, 2019;\cite{schulan19}).

\begin{center}
\textmd{\small DRIVERS' TRAFFIC BEHAVIOUR AND SPEED ENFORCEMENT OF THE POLICE}
\end{center}

Traffic environment is social by nature, a special kind of social situation, in which road users are interacting. Interaction between drivers, who typically are anonymous, is brief and non-recurring. In road traffic, individuals are not united by any traditional group identity, and road users do not know each other by default. Communications are limited in content and are mediated through mechanical aids, such as lights and horns. Driving is inherently a coordinative activity among individuals. It is plausible that trust and norms of trustworthiness would affect the nature of interactions on the roads. (Ross, 1960;\cite{ross} Fleiter et al., 2010;\cite{fleiter10} Tammi, 2013;\cite{tammi13} Nagler, 2013;\cite{nagler13}). Moreover, traffic policing draws police officers into interactions with citizens who may not otherwise receive the attention of the authorities (Schafer and Mastrofski 2005)\cite{schafer05}. The police officer acts as a gatekeeper to the criminal justice system. It is assumed that the officer will act as a 'common sense filter' (Wells 2008, p. 813-814.) \cite{wells08}. Since its early days, traffic enforcement has undermined the relationship between citizens and the police. When the police response to traffic behaviour of a road user and it is perceived as unfair, it can weaken the image of the policing in general, weaken the legitimacy of the police, reduce the citizens' willingness to cooperate and thus weaken the police's chances of succeeding in other tasks that belong to it, which rely on the help, cooperation and active support of the citizens (Huotari, 2020: 140.) \cite{huotari}.

Speed limits are one common rule of the road traffic but not a science for all people or everyday motorists. Why this or that limit? To answer this, you should probably know the statistics, the history of the speed limits on the road, the (serious) crashes that have occurred, the pressure from local residents to lower the limit and the measures taken for one reason or another. The unpopularity of the theory of speed limits and their practice lead to difficulties to follow speed limits rationally from time to time (see Hänti $\&$ Huotari, 2019)\cite{hanti19} Driving faster than the speed limit is globally relatively commonplace (Glendon, 2007;\cite{glendon} Holocher $\&$ Holte, 2019;\cite{holocher}). Speeding is more socially accepted than other traffic violations (see European Commission, 2021:8)\cite{euro}. Though, according to the a Finnish survey speeding drivers arouse the most fear among road users (Liikenneturva, 2023). Many different people are speeders and a majority of drivers admit to speeding at some times (Webster $\&$ Wells, 2000). In a Finnish study, most of the drivers reported that they drive at high speed because of a hurry (Mäkinen, 1990)\cite{makinen}. Being in a hurry largely contribute to drivers exceeding the speed limits, on a general level (Wallén Warner, 2006;\cite{wallen} see also Bogdan et al., 2014;\cite{bogdan}).

Drivers’ speed is influenced e.g. by road factors, other road users and enforcement (Haglund, 2001). Driving at excessive or inappropriate speed is a major threat to safety on the road. It is estimated that up to 15$\%$ of all crashes and 30$\%$ of all fatal crashes are the direct result of speeding or inappropriate speed (European Commission, 2021) \cite{euro}.  Road users are aware of speeding being one main cause of a road crash. Many road users think that accidents due to speeding apply to others, but not to themselves (Holocher $\&$ Holte, 2019) \cite{holocher}. Too high driving speed has been associated with about 40 percent of fatal road accidents investigated by road accident investigation teams in Finland, according to the Crash Data Institute OTI's accident report 2021 (Sihvola 2022)\cite{sihvola}.
 
Criminals are affected by the police, as they avoid committing crimes if the chances of capture are high (Mukhopadhyay et al., 2016)\cite{mukhopadhyay}. Much police-public contact happens on roads which are central to modern lives (Corbett $\&$ Grayson, 2010) \cite{corbett10}. The police follow traffic rules as part of their work (Hänti $\&$ Huotari, 2019)\cite{hanti19}. The police role in traffic has become more extensive and complex as the number of cars on roads and associated traffic offences have very strongly increased, and more complex traffic systems in general (Mäkinen et al., 2003: 53) \cite{makinen03}. The primary role of speed enforcement conducted by the police is deterrence. The police officers, in general, are more interested in preventing speeding than in catching speeders although some drivers believe that the police are out there to get them. As the police develop ever more sophisticated means of apprehending traffic violators, the motoring public develops ever more sophisticated means of countering these efforts. This engages the police and the motoring public in a cat-and-mouse game. If the majority of drivers violate the speed limits, they are very hard to enforce (Shinar, 2017) \cite{shinar}. To keep drivers from speeding, regular police checks are essential. Drivers really often consider it rather unlikely to get checked by the police for speeding on a typical journey (Holocher $\&$ Holte, 2019)\cite{holocher}. 
Jørgensen and Pedersen (2022)\cite{jorgensen05} found that drivers who committed serious speeding offences overestimated the probability of being caught speeding. Older drivers had less knowledge about the threshold level for serious speeding but more knowledge about the detection rate than younger drivers do.

Different traffic violations are not independent of each other but in many cases are associated with each other. Speeding increases the likelihood of other traffic violations, especially of those associated with overtaking (see e.g. Mäkinen et al., 2003) \cite{makinen03}. Research shows a high cross-over between minor and major traffic offending, offending on the roads, and mainstream crime (see Corbett, 2008)\cite{corbett08}. According to Crosetta et al. (2021)\cite{crosetta}, first-time serious traffic offenders are more likely to have a previous or future initial non-traffic offence. Thus, first-time serious traffic offences can act as trigger/strategic offences (Crosetta et al., 2021) \cite{crosetta}. Traffic enforcement is a key way for the police to influence traffic safety and crime in traffic. During enforcement, the police must actively try to uncover non-traffic crimes as well. Crime that occurs in traffic is typically hidden crime that becomes apparent during traffic enforcement (Poliisi, 2022: 14.)\cite{poliisi2022}

Legitimacy, as trust towards the police, influences behaviour strongly. People with higher levels of trust and support for the (law enforcement) authorities are less likely to engage in behaviour against the system (Tyler, 2006: 33.) \cite{tyler}. In traffic, driver beliefs and attitudes are among the strongest predictors of risky driving (Fernandes et al., 2010) \cite{fernandes}. The lack of police enforcement has been found to be a significant predictor of aggressive and risky driving (Stanojević et al., 2018). Favourable attitudes to speeding and experiences of punishment avoidance are among the factors that significantly predict the frequency of speeding (Fleiter $\&$ Watson, 2005)\cite{fleiter05}. However, attitudes and driving behaviour do not always match (Corbett $\&$ Simon, 1999: 82). The attitude and responsibility of each road user contributes to the entire traffic culture. Selfish and aggressive traffic behaviour means more dangerous traffic (Tolvanen, 2018: 44) \cite{tolvanen}. Many young drivers believe that avoiding detection by law enforcement is a breeze (Hoekstra and Wegman, 2011) \cite{hoekstra}. Speed enforcement does not influence the general respect for speed limits (Rintamäki et al., 2022: 73) \cite{rintamaki}. The effectiveness of enforcement is directly related to its perceived threat. To maximize enforcement the driver has to see many police vehicles (that actually seem to control the driving speeds). One effective method to increase the perceived presence is to use a mix of marked and unmarked police vehicles, and make sure through communications that the motoring public is aware of the presence of them (Shinar, 2017) \cite{shinar}. Traffic enforcement is, to a considerable extent, communication on control and campaign related to that. The communication serves educational purposes and recalls police traffic enforcement and the risk of capture (Huotari, 2020: 135.) \cite{huotari}. Visible enforcement influences traffic behaviour and non-visible traffic enforcement increases the experience of the possibility of capture everywhere and at all times (Huotari, 2020: 27) \cite{huotari}. Speeds should be controlled visibly, so that the largest possible number of road users can notice the control. If the police implementate a massive high visibility enforcement operation and announce that, drivers' awareness of the operation has the potential to decrease aggressive driving behaviour. Communication together with high visibility get people to drive more slowly and responsibly to avoid sanctions. (Sonduru Pantangi et al., 2020;\cite{sonduru20} Bauernschuster $\&$ Rekers, 2022.\cite{bauernschuster22}) However, invisible and completely surprising control is also needed to keep the perceived risk of capture as high as possible (Poliisi, 2022: 40) \cite{poliisi2022a}. It is possible to influence road users with different attitudes with wide-ranging and different forms of enforcement (Sisäministeriö, 2021) \cite{sisaministerio}. To influence and to tackle speeding behaviour, police enforcement is one tool but there is a need for multiple intervention tools (see Hankonen, 2019;\cite{hankonen}  Hoekstra and Wegman, 2011;\cite{hoekstra}  Sutela $\&$ Aaltonen, 2021;\cite{sutela}). It has been suggested that high levels of enforcement (about three times the normal level) are needed to obtain effects on drivers' perceived risk of detection (Haglund 2001;\cite{haglund01} Østvik, 1989\cite{ostvik89}).

The effects of speed enforcement on driving behaviour are easily local and short-term, even if capture and being punished quickly follow breaking the law. It would be essential to raise the real and perceived risk of capture. This requires coverage and continuity from the speed enforcement. Generally, there are no resources for this due to the frequency of speed limit violations. The risk of capture is perceived as small indeed (Huotari, 2020: 28) \cite{huotari}.

Ryeng (2012) \cite{ryeng} found that the largest speed reducing effects on individual speed choice were found by either making most other drivers on the road reduce their speed, or by substantially increasing enforcement. Visible police presence in traffic is important, as a measure for reducing average speeds, as well as for the enforcing of minor offences (Vadeby et al., 2018;\cite{vadeby}  see Bogdan et al., 2014;\cite{bogdan} Stanojevic et al., 2018;\cite{stanojevic}). Traffic enforcement creates an objective risk of capture. Control experiences (detection of surveillance and traffic violations) and information on the enforcement in turn create subjective risk of punishment (Mäkinen, 1990)\cite{makinen}. In order to strengthen the effectiveness of traffic enforcement, the available resources must be alloctaed more intelligently and the enforcement must be automated (Huotari, 2020: 26-27) \cite{huotari}. Speed enforcement relies increasingly on automatic traffic enforcement of the police where driving speed are controlled by traffic safety cameras, the advantage of which compared to traditional, manned speed enforfcement is cost efficiency (see Rintamäki et al., 2022: 89)\cite{rintamaki}. The idea that merely seeing the camera poles have an effect on driving behaviour. When drivers see a camera pole, the fear of capture and also other deterrents guide them to drive within legal limits (Sutela $\&$ Aaltonen 2021)\cite{sutela}. 

The police has to prioritise their resources and duties. In helping determine the allocation of police resources, mapping of crime "hot spots" has become an essential tool. It is more useful to consider how police officers are deployed than just increasing numbers of them (see Ratcliffe, 2016: 105, 139)\cite{ratcliffe}. Police effectiveness is a crucial component of policing (Pryce $\&$ Time, 2023)\cite{pryce23}. Traffic enforcement is a traditional, relatively resource-intensive branch of policing (Huotari, 2020: 35) \cite{huotari}. Because the police organisation has a multiple goal structure, resource deployment is a complex task, since effectiveness in one area may mean ineffectiveness in another (Chatterton 1987, 81) \cite{chatterton}. In the police organisations, traffic safety work easily appears as a secondary task, which can generally be moved aside without necessarily influencing anyone, but the resources used or attached to it are always away from other (actual or proper) police work (Huotari, 2020: 131) \cite{huotari}. The resource deployment strategies of the police are often decided in an ad-hoc manner (Samanta et al., 2022). Police departments decide quite independently how traffic enforcement is organized in the area and how its resources are determined and distributed among different functions of the unit (Karnaranta, 2019: 51) \cite{karnaranta}. 

Also traffic enforcement needs prioritisation. Forms of misconduct in traffic behaviour are numerous and focusing on them all systematically in enforcement is simply impossible. Relatively few drivers may be influenced, only a few aspects of driving behaviour may be focused on. There are also reasons other than resources why the police need to set priorities for enforcement. These are associated with the capabilities of the police to measure and detect various violations (Mäkinen et al., 2003: 48.) \cite{makinen}. As a response to crime citizens strongly support aggressive traffic enforcement practices focusing on "hot spots". However, it has been suggested that law enforcement officials often develop and implement strategies without public input (Chermak, McGarrell $\&$ Weiss, 2001)\cite{chermak}. Communication of enforcement strategies and themes should be regarded as an important part of the effectiveness of enforcement. The citizens should be involved in target setting and actively asked to contribute to the achievement of the traffic safety objectives (see Mäkinen et al., 2003: 79.)\cite{makinen03}.

Game theory deals with any situation in which a strategy i.e. is a plan for acting that responds to the reactions of others is important. To characterize a game, we must specify three things: players, strategies of each player, and payoffs to each player for each strategy (Cooter $\&$ Ulen, 2012: 33.)\cite{cooter12}. In Bjørnskau and Elvik (1992)\cite{elvik0} it is suggested that game theory has been shown to be \emph{most fruitful} in analysing police and road user interaction with respect to traffic law violations. Essentially, this line of argumentation is based on the idea that the traffic can be seen as a game where the drivers gain utility from speeding and the police from possibility to allocate resources to some other policing operations than traffic enforcement. Given that road traffic is a very rich phenomenon, as already established in this article, it's very clear that no mathematical model of ours will be able to encompass everything that factors into the road user dynamics. Thus, we will focus our attention to incentives to violate the speed limits and incentives to prevent such violations with an abstraction of the strategic interaction between the police and the drivers insipred by earlier research and the more complicated structure that can be generated from that basis mathematically. Everything beyond the game theoretic model consists of \emph{externalities} that are maily discussed only if they bear any relevance (real-life or theoretical) to what emerges from that analysis.

A central and famous concept of game theory is \emph{Nash equilibrium} -  named after American mathematician John Nash (1928-2015) - for which there exist many refinements. In this article, we wish to examine whether the game played by the drivers and the police has such an equilibrium. We will do this mainly by following and elaborating on Elvik (2015)\cite{elvik}. Elvik explored the relationship between changes in the risk of apprehension for speeding in Norway
and changes in the amount of speeding. Unlike Elvik, we do not attempt to test the game-theoretic model empirically. The Norwegian basis of our paper can be justified by the fact that Finland and Norway are very similar countries, for example in terms of main characteristics of mobility (Pöllänen et al., 2023)\cite{pollanen23}. Norway is one of the best-performing countries for road safety. For the improvement for road safety, Finland can learn from Norway by e.g. lowering speeds and enhancing compliance with speed limits (Pöllänen et al., 2023)\cite{pollanen23}. As Elvik, we focus too on speeding violations, due to speeding has one of the highest priority in traffic policing (Mäkinen et al., 2003: 56)\cite{makinen} and because, as stated above, speeding is such a traffic violation drivers gain utility from.

\begin{center}
\textmd{\small A SIMPLE SEQUENTIAL GAME MODEL AND NASH EQUILIBRIA}
\end{center}
 
First of all, it's important to note that a Nash equilibrium is not "good" or "bad" by its plain definition. Essentially it just implies that all players' stragies are best responses to each other and no party has an incentive to deviate by changing one's strategy. Every finite game will always have at least one equilibrium, which was shown by Nash in 1950\footnote{This was proven by Nash in his paper \emph{Equilibrium points in \texttt{n}-person games} (January 1950) when working at Princeton University.}. This equilibrium may be in \emph{pure} or in \emph{mixed} strategies. Here, "pure" refers to strategies or actions being played with a probability $1$ or $0$, whereas "mixed" refers to strategies or actions being played with probabilities from the open interval $(0,1)$. A mixed strategy Nash equilibrium, however, may be a bit abstract and difficult to interpret in our setting of drivers and traffic enforcement of the police. The best interpretation seems that drivers should violate and certain \emph{frequency} and the police should enforce at another frequency.

\setcounter{footnote}{0}
 
\begin{table}[H]
\centering
   \setlength{\extrarowheight}{2pt}
   \begin{tabular}{cc|c|c|}
     & \multicolumn{1}{c}{} & \multicolumn{2}{c}{Drivers}\\
     & \multicolumn{1}{c}{} & \multicolumn{1}{c}{$S$}  & \multicolumn{1}{c}{$DS$} \\\cline{3-4}
     \multirow{2}*{Police}  & $E$ & $-10000,-300$ & $-10000,-50$ \\\cline{3-4}
     & $DE$ & $-20000,50$ & $0,-50$ \\\cline{3-4}
   \end{tabular}
\caption{\small{\small{Normal form game matrix from Elvik (2015) [\!\!\!\citenum{elvik}].}}} \label{table:elvikpeli} 
\end{table}
 
What may also seem particularly problematic at first is the idea that the drivers should make choices collectively, as a kind of a rational "hive mind". However, earlier research on the topic provided by Bjørnskau and Elvik (1992)\cite{elvik0} argues that drivers should not be seen as 

\begin{center} 
\emph{passive (parametric) receivers of safety measures} 
\end{center} 

\noindent but rather as 

\begin{center} 
\emph{strategic actors who take advantage of whatever measure available to attain} their \emph{goals} 
\end{center}
 
\noindent (with emphasis on "their") regardless of whether it makes the roads safer or not. In the paper it is further argued that considering the drivers as parametric receivers is in fact a \emph{fatal mistake} potentially resulting in \emph{no-safety effect of potential safety measures}. Watson et al. (2015) \cite{watson} has suggested speeding offenders are not a homogeneous group: important differences exist between different types of offenders, based on the extent and severity of their offending behaviour.

Certainly it always makes sense to suggest that every individual driver seeks to act according to his own interests and that ultimately a driver may be (and even should be), concerned about the utility of the group to which he belongs. In all scenarios the driver's utility with respect to his or her speeding behaviour is dependent on the enforcement strategy of the police. Rather than trying to predict the outcome of this strategic interaction from some equilibria, the key approach here is to examine whether, and under which circumstances, an individual driver has an \emph{incentive} to violate speed limits and if there is a way to \emph{diminish} such incentives.
 
Note that simply having incentives to perform certain actions does not imply that an individual automatically acts according to those incentives. The individual agent may not even be aware of such incentives in the first place. Therefore, rather than relying on every driver computing his or her expected utility every time upon entering the vehicle to find a mathematically optimal strategy, the focus is ultimately on controlling and restricting the road users' potentially utility-maximizing behaviour by an optimal set of rules (or a "mechanism") that has a minimal amount of societally harmful loopholes that calculative law-breakers might take advantage of.

Traffic enforcement of the police is largerly revealing activity. The more the police control, the larger number of traffic crimes and violations are uncovered (Kautto, 2021) \cite{kautto}. Periodic traffic enforcement i.e. intermittently enforcement throughout a specific time period has been found to be an effective in reducing average speed and the proportion of drivers exceeding the speed limit. The existence of such law enforcement could potentially modify drivers' behaviour. (Karimpour et al., 2021)\cite{karimpour21}. Elvik (2015)\cite{elvik} points out that the police adapts to changes in violation periodically. The existence of such periods implies that a model consisting of a \emph{repeated game} may be appropriate. On the other hand, even the notion of the police adapting to something the drivers are doing implies that the game has to be \emph{sequential} and not simultaneous. Potential offenders, including traffic offenders, are rather aware of changes in policing. There is considerable and robust evidence that crime, also traffic crime, is responsive to police operations, increases in police manpower and to many varieties of police redeployments. Individuals respond to the incentives that are the most immediate and salient (see Chalfin $\&$ McCrary 2017)\cite{chalfin}. The police would move first and annually (or quarterly or even monthly) after learning information about the traffic violations of the past year (or some previous weeks or months). So rather than starting by postulating a simultanous-move normal form game that is played once, perhaps a sequential, infinitely\footnote{Clearly, ''infinitely'' here is just an abstaction referring to the idea that a game is repeated for a very long time or even that it's is only played for a short or medium duration but consists of a countless number of periods. We will discuss period durations later in this paper in more detail.} and periodically repeated game would model better the dynamics at hand. We'll start by modeling an extensive form game where the police moves first and the drivers second\footnote{What if, hypthetically, we reversed the setting and let the drivers move first? The result is that we get a game with a different pure strategy SPE through backward induction. In fact, in this SPE the police are not enforcing and the drivers are not speeding - exactly the kind of equilibrium the police would prefer. If the police were to mix in the off-path node, the drivers would have an incentive to switch their strategy given that $\mathds{U}_D(S) > \mathds{U}_D(DS) \Leftrightarrow b_P(E)<1/4$ where $\mathds{U}$ denotes expected utility (ommitting the steps due to the similarity to calculations on p. \pageref{eqn:mix1}).\par Would it, however, make sense to let the drivers move first when modeling a real-life scenario? In the game-theoretic model the advantages and disadvantages of speeding stem solely from the enforcing strategy of the police. Thus, it's not too far fetched to postulate that a driver would not be prevented from speeding by the idea that his criminal behaviour would have a minor influence to the enforcement strategy of the police next year. In order to directly appeal to his incentives, we should have an enforcement strategy in place while the person is making his speeding decision. This is why we choose to stick to the model where the police is moving first, or at least \emph{not later than the drivers}, every period.}.

\tikzset{
solid node/.style={circle,draw,inner sep=1.5,fill=black}, 
hollow node/.style={circle,draw,inner sep=1.5} } 

\begin{figure}
\begin{center}
\begin{tikzpicture}[scale=1.2,font=\footnotesize]  
\tikzstyle{level 1}=[level distance=15mm,sibling distance=35mm] 
\tikzstyle{level 2}=[level distance=15mm,sibling distance=15mm] 
\node(0)[solid node,label=above:{Police}]{} child{node(1)[solid node, label=right:{Drivers}]{} child{node[hollow node,label=below:{$(-2\cdot 10^4, 50)$}]{} edge from parent node[left]{\emph{Speed}}} child{node[hollow node,label=below:{$(0,-50)$}]{} edge from parent node[right]{\emph{Don't Speed}}} edge from parent node[left,xshift=-3]{\emph{Don't Enforce}} } child{node(2)[solid node, label=right:{Drivers}]{} child{node[hollow node,label=below:{$(-10^4,-300)$}]{} edge from parent node[left]{\emph{Speed}}} child{node[hollow node,label=below:{$(-10^4,-50)$}]{} edge from parent node[right]{\emph{Don't Speed}}} edge from parent node[right,xshift=3]{\emph{Enforce}} }; 
\end{tikzpicture}
\end{center}
\caption{\small{Extensive form for the sequential game with the police moving first.}} 
\end{figure}
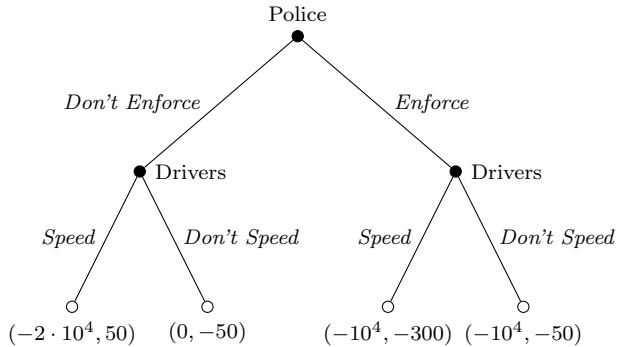

Now, this extensive form game is different from the original normal form game (as seen in Table \ref{table:elvikpeli}) and it has a Nash equilibrium in \emph{pure strategies} where the police are enforcing traffic and the drivers are not speeding on the path of play. This equilibrium can be found through a process called \emph{backward induction}. If the driver finds himself in the left decision node, he would like to play "Speed". In the right decision node, he would like to play "Don't Speed". Proceeding backwards, the police would then like to play "Enforce" for a bigger payoff. Therefore, we have found a pure strategy\footnote{A common style of denoting player $i$'s pure strategy is $s_i$ which is a member of that player's set of pure strategies $S_i$, i.e. $s_i \in S_i$. In this case, given that P is short for Police and D short for Driver(s), the SPE profile (\ref{eqn:spe1}) is $(s_P, s_D)$, where $s_P$ equals \textrm{Enforce} and $s_D$ equals (\textrm{Speed (in the left node), \textrm{Don't speed (in the right node)}}). } Nash equilibrium of the game
 
\begin{equation}
\label{eqn:spe1}
\centering
(\textrm{Enforce}, (\textrm{Speed (in the left node)}, \textrm{ Don't speed (in the right node)}) 
\end{equation}
 
This Nash equilibrium is in fact called a \emph{subgame perfect equilibrium} (SPE), which means that this strategy profile corresponds to a Nash equilibrium in every \emph{subgame} of the full game. This game has three subgames, the two starting at the driver's decision nodes and the full game (a game is always a subgame of itself). Now that we've found the SPE of the game, we know that the driver will not have an incentive to deviate from not speeding when the police is controlling and also the police will not have an incentive to switch from control to not control when the drivers are not speeding.\par Even though this is an SPE, it is not an outcome the police would like. In fact, they would naturally prefer the outcome where "Don't Enforce" and "Don't Speed" are played on the path of play, which is the outcome we would have in a world without speeding. Since this outcome would yield the drivers exactly what they obtain in the SPE outcome, this outcome is in fact a \emph{Pareto improvement}, i.e. nobody's utility is taken away but at least someone is better off. Nevertheless, no pure strategy equilibrium where the police plays "Don't Enforce" and the drivers play "Don't Speed" exists since knowing that the police are not enforcing the drivers will have an incentive to change their strategy to speeding on the path of play.\par
 
We should examine off-path behaviour that involves \emph{mixed strategies}. A mixed strategy is simply a probability distribution over the set of pure strategies. This means that in every node the sum of the probabilites of the possible actions is always one, i.e. $100\%$. The drivers are rational to mix only if they are indifferent between their choices of pure strategies or in decision nodes that are off the path of play. Notice that, as derived in Bjørnskau and Elvik (1992)\cite{elvik0}, there exists a mixed strategy equilibrium in the simultaneous-move game as given in Table \ref{table:elvikpeli}. In this equilibrium, the drivers are speeding 50 percent of the time and the police is enforcing a bit less than 30 percent of the time. However, the authors do not find this equilibrium ``stable'' due to the enforcer reducing the level of enforcement whenever the number of traffic law violations is decreasing. This implies there clearly exist some externalities to the game as given by Table \ref{table:elvikpeli} since the police is clearly expecting the drivers to change their strategy soon in the future - something that isn't captured by the simple $2\times2$ game matrix alone.

To discuss mixing further with respect to the sequential form game, assume that the drivers are mixing between speeding and not speeding in their left decision node. The police then has an incentive to switch from enforcing to not enforcing given that their \emph{expected utility}, simply the expected value of their utility, from not enforcing with respect to the drivers' mixed strategy is greater than from enforcing. We denote the drivers mixed strategy $$(b_D(\textrm{Speed (in the left node)}), b_D(\textrm{Don't speed (in the left node)})),$$ where "b" refers to a \emph{behaviour strategy} (simply a name for a mixed strategy in a sequential game). To simplify our notation a little bit, we now denote the action of speeding in the left node by $S_{L}$ and not speeding in the left node by $DS_{L}$. We also shorten the notation for other strategies similarly: enforce will be $E$ and don't enforce $DE$, with an appropriate subscript ("$L$" for an action played in the left subame and "$R$" in the right) if needed. Then it's possible to write the aforementioned behaviour strategy as $(b_D(S_{L}), b_D(DS_{L})) = (b_D(S_{L}), 1-b_D(S_{L}))$. Let us denote the function for expected utility with the symbol $\mathds{U}$ with an appropriate subscript. In the case of the drivers using a mixed strategy in the left node, the expected utility of the police is

\begin{align}
\label{eqn:mix1}
\mathds{U}_P(DE) &= -2 \cdot 10^4b_{D}(S_L) + 0 \cdot b_{D}(DS_L)\\
					  & = -2 \cdot 10^4b_{D}(S_L)\\
\vspace{1em}
\shortintertext{Therefore the police plays "Enforce" as long as}\\
\vspace{1em}
\mathds{U}_P(DE)&< \mathds{U}_P(E)\\
\label{eqn:mix2}
\Leftrightarrow -2 \cdot 10^4 b_{D}(S_L) &< -10^4\\
\Leftrightarrow \hspace{1em}b_D(S_L) &> 1/2
\end{align}

Then
 
\begin{equation}
\label{eqn: spe2}
\centering
\{(E, ((b_D(S_L), 1-b_D(S_L)), DS_R) : b_D(S_L) \in (1/2, 1) \} 
\end{equation}

\noindent is a Nash equilibrium (or a spectrum of equilibria given by varying $b_D(S_L)$) of the game. However, this is not an SPE since mixing in the left subgame is not a Nash equilibrium of that subgame. \par
 
How would this translate into a real-life setting, if at all? Exceeding the speed limit may often be a deliberate choice. Speeding can be defined as a type of driver error, deliberate error i.e. violation (Haglund, 2001: 13.)\cite{haglund01}. People drive deliberately at excessive speed out of habit (De Pelsmacker and Janssens, 2007) \cite{depelsmacker}. Over half (57$\%$) of drivers who responded to the survey by the Finnish Road Safety Council (Liikenneturva, 2020)\cite{liikenneturva} told that slight excess of speed limits has become a habit. In 2020, on Finland's main roads, about 43 percent of drivers drove over the speed limits in summer period and 54 percent in winter period (Kiiskilä, Mäki $\&$ Saastamoinen 2021) \cite{kiiskila}. From the perspective of the police, drivers speeding with a probability of less than half could be understood as less than half of the large population of drivers speeding and the rest not speeding. Knowing that the drivers are speeding with a probability greater than 50 percent, the police would not have an incentive to deviate from controlling. Knowing that the police are controlling, the drivers as a group would not have an incentive to speed.\par Deterrence is an old idea and has been discussed in academic literature at least from eighteenth-century. There are three core concepts embedded in theories of deterrence. According to them, individuals respond to changes in the certainty, severity, and celerity (or immediacy) of punishment (see Chalfin $\&$ McCrary, 2017) \cite{chalfin}. The deterrence hypothesis suggests that individuals balance off the costs and benefits of crime in deciding whether to engage in criminal activity. According to the the deterrence approach, following Becker (1968), increasing the probability of capture (and punishment), for example by increasing the number of police officers, should deter criminal behaviour (Dnes, 2000: 72–73.\cite{dnes00}). Control experiences and information on the enforcement should create subjective risk of capture. Influencing and increasing drivers´ risk perception i.e. subjective risk of capture is one of the most significant factors to prevent traffic violations, such as speeding. However, it has quite a marginal effect to drivers who repeatedly are committing traffic violations (e.g. Mäkinen 1990 \cite{makinen};  Tolvanen 1999;\cite{tolvanen}). Police enforce road rules through strategies that aim to create a credible risk of punishment for wrong-doing. However, more recent theorising has proposed that informal deterrence, instead of formal one, is most likely to influence behaviour. Researchers have claimed that those who have significant relations and social bonds to would-be offenders aremore likely to have success in deterring their non-compliance because they are more likely to elicit internal feelings of shame over wrong-doing among the recipient. For example, parents have to be found to deter young drivers non-compliance more likely than police (Allen et al., 2017;\cite{allen17} Murphy $\&$ Helmer, 2013;\cite{murphy13} Grasmick and Bursik, 1990;\cite{grasmick90}).
 
However, if $b_D(S_L) \leq 1/2$, i.e. the drivers are speeding with a probability no more than 50 percent, the police would have an incentive to deviate to not control. Notice that (\ref{eqn:mix2}) tells us that the only way to alter the treshold of $b_D(S_L)$ this NE is to alter the ratio $$\frac{u_P(E, DS_R)}{u_P(DE,S_L)} = -2 \cdot 10^4 / 10^4$$ which is independent of the gains and penalties of the drivers. This could mean, for example, that the police are not going to change their enforcement strategy before the drivers are speeding with the probability of less than 80 percent or 30 percent or 15 percent... all depending on the ratio of the utility of controlling while the drivers are not speeding and not enforcing while the drivers are speeding. \par

\begin{center}
\textmd{\small THE PHENOMENON OF ALTERNATING STRATEGIES}
\end{center}
 
According to Elvik (2015)\cite{elvik} the violations will start to drop when the police are controlling traffic and rise when it's not controlling. We discuss and try to model this "oscillating" behaviour mathematically by using our simple sequential game model as a stage game for an indefinitely (not necessarily infinitely) repeated game. We also try capture and include the behaviour of the drivers as a group, which is ultimately to which the police responds with a choice of strategy. In a repeated game this aggregate behaviour could be modelled with an appropriately decreasing or increasing sequence $\{ b_D^t (S_L \vline E) \}_{t \in \mathds{N}}$, where the superscript $t$ is the number of the period. Here, $S_L$ and $E$ separated by a vertical line means \emph{the probability of $S_L$ conditional on $E$}. However, we have to careful here. Since we are addressing speeding in the left subgame, we are talking about the drivers' choice of action as a response to the police not enforcing while on the path of play the police are still enforcing and the drivers not speeding. Game theoretically this is to think of what the police think and know that the drivers currently \emph{would do} if the police \emph{were not} controlling. Since on the path of play the drivers are not speeding, this means they are playing "don't speed" with probability one (also called a \emph{degenerate} mixed strategy). Again, rather than thinking about this strategy as "most" drivers not speeding, it's more appealing to think of it as a pure strategy of an individual driver. The behaviour strategy in the left node, however, can be understood as a probability distribution over an individual driver's pure strategies and more interestingly, as the statistical distribution of the actions taken by the population of drivers. The police are not going to switch from enforcing to not enforcing unless the population's response to that is to not speed with the probability of one or close enough to one. The phenomenon of the violations decreasing during the police enfocing traffic is therefore consistent with an individual driver varying his or her behaviour strategy in the left subgame such that $ b_D^t (S_L) $ is drawn from a decreasing sequence. Once $ b_D^t (S_L) $ reaches $1/2$, the police will switch to not enforcing. As the violations will begin to rise once the police is not enforcing, $\{ b_D^t (S_L) \}_{t \in \mathds{N}}$ will continue as an increasing sequence $\{ b_D^t (S_L \vline DE) \}_{t \in \mathds{N}}$.\par
 
If $ b_D^t (S_L) $ drops to $1/2$\footnote{This means $ b_D^t (S_L) \geq 1/2$ and $ b_D^t (S_L) \downarrow 1/2$ while $t$ is increasing.}, the drivers are acting against their best interest because they don't have a rational incentive to change their Nash equlibrium strategy in our stage game. This means the drivers are acting irrationally as a group in our game theoretic model. On the other hand, they are acting not only with lawful intentions, but also increasingly many drivers are acting rationally with respect to their own utility in a \emph{myopic} sense (for each driver it's within his or her best interest to not to speed while the police are enforcing, even if the police would change their enforcement strategy some time in the future). In this scenario, the drivers changing their mixed strategy in the left subgame can be seen as an "overcorrection" to their behaviour with respect to the current traffic enforcement. Generally speaking, there is no "wrong" or "right" way to model irrational behaviour, but in our case we can draw the pattern of this behaviour from what has been observed. This is also consistent information incompleteness of a real-life setting: the drivers would not know what the exact limits of the probability (or frequency) interval that keeps the strategy of the police unchanged. Rather, the drivers are getting a ''feel' for the limits.  Also, people tend to follow the herd behaviour in traffic. Drivers tend to influence one another which is nicely reflected in the finding that fast drivers, as compared with slow drivers, thought a higher number of other drivers were speeding. It is much easier to see the consequences of an action, such as speeding, if you watch someone else do it first. Sometimes the instinct to follow the crowd, can get us into trouble (Orrell, 2021: 130;\cite{orrell21} Haglund, 2001: 42\cite{haglund01}).

Nevertheless, this behaviour gives the police an opportunity to get out of the SPE where they are enforcing while the drivers are rational not to speed. Interestingly enough, once $ b_D^t (S_L) = 1/2$, the police would have the chance to meet the drivers at the unique mixed strategy equilibrium with the best response of enforcing around 30 percent of the time\footnote{The exact mixed strategy equilibrium is $((\frac{1}{2}, \frac{1}{2}),(\frac{2}{7},\frac{5}{7}))$ as derived in Bjørnskau $\&$ Elvik (1992)  [\!\!\!\citenum{elvik0}] .}.  But because the drivers are speeding less and less - to which the police are also getting a ``feel'' by the same figure of speech as before - the police anticipates that drivers speeding probability is not going to remain fixed at $50$ percent. Thus, there is no stable equilibrium at hand, as argued by Bjørnskau $\&$ Elvik (1992)\cite{elvik0}. Once the police stops enforcing, the each driver is immediately rational to play the pure strategy "Speed" to which the police would again be rational to respond "Enforce". However, even though an individual driver gains the incentive to start speeding, in the aggregate the drivers can't be expected to change their strategy immediately but rather follow the sequence of behaviour strategies where $b_D^t(S_L \vline DE)$ is increasing.
\par

\tikzset{
solid node/.style={circle,draw,inner sep=1.5,fill=black}, 
hollow node/.style={circle,draw,inner sep=1.5} } 

\begin{figure}
\begin{center}
\begin{tikzpicture}[scale=1.2,font=\footnotesize]  
\tikzstyle{level 1}=[level distance=15mm,sibling distance=35mm] 
\tikzstyle{level 2}=[level distance=15mm,sibling distance=15mm]  
\node(0)[solid node,label=above:{Drivers}]{} child{node(1)[solid node, label=right:{Police}]{} child{node[hollow node,label=below:{$(-50,-10^4)$}]{} edge from parent node[left]{\emph{Enforce}}} child{node[hollow node,label=below:{$(-50,0)$}]{} edge from parent node[right]{\emph{Don't Enf.}}} edge from parent node[left,xshift=-3]{\emph{Don't Speed}} } child{node(2)[solid node, label=right:{Police}]{} child{node[hollow node,label=below:{$(-300,-10^4)$}]{} edge from parent node[left]{\emph{Enforce}}} child{node[hollow node,label=below:{$(50,-2\cdot 10^4)$}]{} edge from parent node[right]{\emph{Don't Enf.}}} edge from parent node[right,xshift=3]{\emph{Speed}} }; 
\end{tikzpicture}
\end{center}
\caption{\small{Extensive form for the sequential game with the drivers moving first (see f.n. on p. 7)}} 
\end{figure}
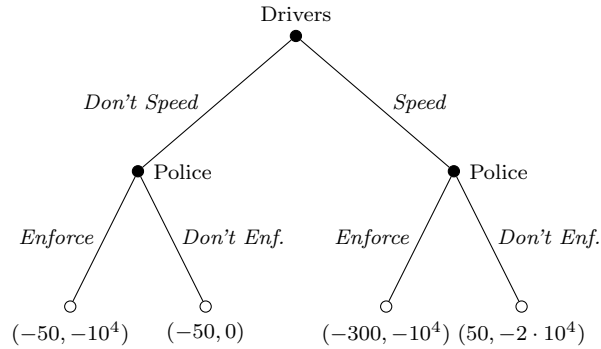

So far we haven't discussed too much about information. The reason we have obtained such an SPE in the game-theoretic model we have developed so far, is that we have assumed the game to be of \emph{perfect information}. This means that all players know each others' stragegies, utilites and past and future play. They know what each player does (or "would do") in each subgame. Notice that the notion of drivers having incentives and being aware of them relies on this kind of information assumptions.
 
How realistic is this assumption with respect to a real-life setting? The police obtains statistical information about drivers' violations from previous periods, which probably is in line with this perfect information assumption farely well. However, it cannot be completely clear to the police what happens \emph{in the future} after they make their choice of strategy. When it comes to the drivers, this assumption loses its credibility even more. It is not too unrealistic to postulate that in the aggregate the drivers are not completely certain of whether the police are enforcing traffic or not. On the other hand, the idea of an incentive to deviate from not speeding to speeding hinges on the notion of being aware of the "opponent's" current strategy. To simplify things, however, we choose to take into account only the drivers who are well aware of the actions of the police and model them as having perfect information. Better yet, we can equivalently view the situation from the point of view of an individual driver who has perfect information and plays the game alone against the law enforcement. 
 
What happens if this stage game is repeated infinitely? Would playing (\ref{eqn: spe2}) or  (\ref{eqn:spe1}) infinitely also be an SPE of this game? To answer this question, we first recall that (\ref{eqn: spe2}) itself isn't an SPE profile, which directly rules it out. Profile (\ref{eqn:spe1}), however, is an SPE for the game in a repeated form given any number of periods\footnote{Suppose the game is played for $T$ periods. If the profile isn't an SPE, there would exist some $t\leq T$ such that on $t$ one of the players has an incentive to deviate from following the profile. But this is impossible since a NE is played on $t$.}. This begs the question: how is this information \emph{useful} in any way? Clearly, it would serve as an odd prediction of the play in most societies. It's also a very undesirable and unrealistic outcome for both parties in most scenarios. The actions (or \emph{action profile}) played on the path of play in (\ref{eqn:spe1}) (and in  (\ref{eqn: spe2})), first \emph{Enforce} and then \emph{Don't Speed}, do not constitute a NE of the plain simultaneous-move stage game, which mostly takes away the predictive power of the outcome . The equilibrium only arises once the game becomes sequential and the off-path actions for both players common knowledge. This shows how abundant information is absolutely crucial for the existence and relevance of this kind of equilibria and how an action profile that would not be an equilibrium in itself can be sustained as an \emph{equilibrium outcome} given the structure of the entire game\footnote{An archetypical example of this is the \emph{Prisoner's Dilemma} where a profile consisting of repeated play of actions otherwise strictly dominated in the stage-game become a supportable outcome only in the case of infinite periods. See Mailath (2018)\cite{mailath}, chapter 7.}. As we move on to repeated-game-framework, we'll see how the game being endowed with new structure could in principle allow us to sustain a wide variety of outcomes in the equilibrium - some of which may be worthwhile to consider with respect to a real-world setting. 

\begin{center}
\textmd{\small INFINITELY REPEATED GAME AND ITS FORMALISM}
\end{center}
 
We move on to examine whether the repeated game framework allows us to construct feasible SPE strategy profiles with respect to our game theoric setting derived from Elvik (2015) \cite{elvik}. We are especially interested in SPEa that allow the police to escape enforcing every period that was implied in the earlier analysis with the sequential form stage game. Also, our focus shifts from just modeling the phenomena game theoretically to trying to solve optimal policing strategies in a \emph{carrot-and-stick} fashion. Doing this, we avoid going too deep into the general theory of repeated games - for which a  great resource is Fudenberg and Tirole (1991)\cite{fudenbergtirole}. In repeated games, one must consider an economic factor called the \emph{discount factor}, commonly denoted by $\delta$, which represents the agent's devaluation of the good after one perior. For example, if $\delta = 1$ the agent's valuation of the good is unchanged by a period change, and if $\delta = 0$, the good has no value at all to the agent on the following period. In this article we make no assumption that the discount factor of the drivers should differ from the one of the police.
 
Consider constructing an automaton with stage-game utilites given by the 2$\times$2 game matrix as given in Elvik (2015)\cite{elvik}. Denoting an action profile for period $t$ as $a(t) = (a_D(t),a_P(t))$, the utility for repeating an infinite sequence $\{a(t)\}_{t\in \mathds{N}}$ for a player is
 
\begin{equation}
\label{geom_sum}
(1-\delta)\sum_{t\in \mathds{N}} \delta ^{t-1}u_i(a(t)) 
\end{equation}
 
\noindent where $i \in \{ D,P\}$. Since we are discussing \emph{infinitely} repeated games, certain interesting dynamics come into play that set them apart from finitely repeated games. Clearly, a backwards induction cannot be applied to these games since there is no final period. Nevertheless, it's possible to find SPEa using \emph{automata}. An automaton is a tuple $(\mathcal{W}, w^0, f, \tau)$ such that $\mathcal{W}$ is the set of \emph{states} with an initial state $\omega^0$, $f: \mathcal{W} \rightarrow  A$ is an output function and $\tau : \mathcal{W} \times A \rightarrow \mathcal{W}$ is the transition rule. For a given a state $w \in \mathcal{W}$, we have that $f(w)$ is the set of actions the players are expected to play in that state given that they're following the automaton and $\tau(\omega,a)$ is the state the players are going to transfer to after whichever action profile $a$ they play.
 
For any state $\omega\in\mathcal{W}$ in the automaton and for any player $i\in \{ D, P\}$ there is an associated value $V^{i}_{\omega}$. Intuitively, this is player $i$'s "value for being in the state $\omega$". These state-wise values become critical when investigating whether an automaton describes an SPE of the infinitely repeated game. A necessary condition of an automaton describing an SPE is that \emph{no player} has an incentive to deviate (that is, playing actions other than prescribed by the automaton) from \emph{any} state of the automaton. Specifically, a deviation refers to a \emph{one-shot deviation} where we only assume a deviation from the automaton that takes place on a single period.
 
\begin{theorem}
A strategy profile given by an automaton $(\mathcal{W}, w^0, f, \tau)$ is an SPE if and only if no player has profitable one-shot deviations from any state $\omega \in \mathcal{W}$.
\end{theorem}
 
A proof can be found in Mailath (2018)\cite{mailath}, chapter 7. With respect to the notion of a one-shot deviation, we must define a suitable definition for one period. With respect to the period duration, we start by following Elvik (2015)\cite{elvik} where it is stated that the police collects the data from violation annually and chooses their strategy accordingly.\footnote{Ultimately, however, it becomes very obvious that the period duration does not really make a difference within the model when it comes to optimal strategies, even though it obviously would in reality due to the gains and losses for both parties getting higher. This is because the utilities for the players will away be multiples of the stage-game utilities such that the multipliers will always cancel out of the inequalities for comparing total utilities. To illustrate, when the drivers not speeding and the police are enforcing for one year, the resultant utility the utility vector ($-10000, -50$) multiplied by some $\alpha\in\mathds{R}_{+}\setminus\{0\}$, i.e.  $\alpha(-10000, -50)$ such that $\alpha$ will cancel out. This will be discussed in detail later in the article.} 
 
In our model, the police has no other \emph{endogeneous} way of influencing the drivers' incentives to speed other than to alter the level of traffic enforcement. This can be viewed as \emph{rewarding} and \emph{punishing} the drivers - that is - the police punishes the drivers due to their past speeding by enforcing and rewards them by not enforcing. An essential research question is: if the drivers should and can be punished, how long should the punisment last and will the police have an incentive to carry out the punishment.
 
It seems the only sensical way of the police to punish the drivers for speeding is to use one period, i.e. one year, for enforcement in case the drivers have been speeding excessively. This means that in the case of too much traffic violations, the automaton should be such that the drivers are taken on a \emph{punishment path} where they are punished for one period (year) and one period only. We try to construct a two-state automaton describing an SPE such that the initial state $\omega_0 \in \mathcal{W}$ is the societally optimal one with no player having an incentive to deviate from playing $f(\omega_0)$ due to threat of being taken on a punishment path. Given the $2 \times 2$-stage game matrix from Elvik (2015)\cite{elvik}, it's tempting to try to construct an SPE automaton such that the initial state is $\omega_{(NE,S)}$ and the punishment state is $\omega_{(E, DS)}$. Obviously, the punishment state cannot be $\omega_{(E,S)}$ since - assuming perfect information and that the police moves first on each period - the drivers will not speed knowing that the police is controlling. However, in this automaton the drivers are never rewarded with being able to speed and the stage-game utility with respect to the punishment state is equal to the initial state (i.e. there is no "punishment" at all) and the drivers will always have an incentive to play $S$. It's then clear that an infinite repetition of the play $(DE,DS)$ cannot be sustained.

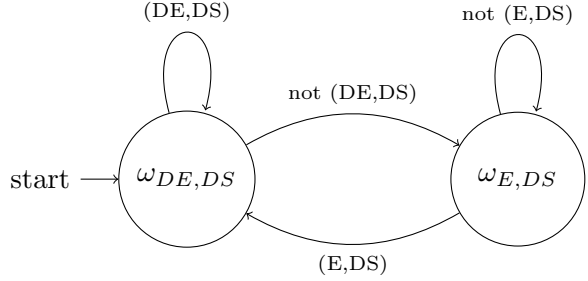
\begin{figure}
\begin{center}
\begin{tikzpicture}[scale=1.1, every node/.style={scale=1.1}]
\node[state, initial] (q2) {$\hspace{0.2em}\omega_{\tiny{DE,DS}}\hspace{0.2em}$};
\node[state, right of=q2, xshift=3cm] (q3) {$\hspace{0.5em}\omega_{\scriptsize E,DS}\hspace{0.5em}$};
\draw[->] (q2) edge[loop above] node{\scriptsize (DE,DS)} (q2);
\draw[->] (q2) edge[bend left, above] node{\scriptsize not (DE,DS)}(q3);
\draw[->] (q3) edge[bend left, below] node{\scriptsize (E,DS)}(q2);
\draw[->] (q3) edge[loop above] node{\scriptsize not (E,DS)} (q3);
\end{tikzpicture}
\end{center}
\caption{\small{Automaton i This fails to be an SPE.}}
\label{auto1}
\end{figure}

Even though this failure of an SPE is quite obvious, we can also show this formally with calculations and more importantly introduce mathematical apparatus generally needed for this kind of analysis. For this, consider the following payoff function (see e.g. Mailath (2018)\cite{mailath}):
 
\begin{definition}{(Payoff function)}
Given a player $i\in I$ and a transition function $\tau : \mathcal{W} \times A \rightarrow \mathcal{W}$ we define a \emph{payoff function} $g_i^{\omega}: A\rightarrow \mathds{R}$ as
 
$$g_i^{\omega} (a) = (1-\delta )u_i(a) + \delta V^{i}_{\tau(\omega, a)}.$$
\end{definition}

Player $i$'s payoff function $g^{\omega}_i$ thus takes an action profile $a\in A$ and a state $\omega \in \mathcal{W}$ as parameters. It returns the sum of the utility of the action profile $a$ being played for one round and the discounted value of the state that the players are being transferred to by the automaton according to the transfer rule $\tau$ given that the players are initially in state $\omega$ and an action profile $a$ is played.
To then show that the automaton in Figure $\ref{auto1}$ fails to be an SPE, we note that in order for the drivers to have no incentive to deviate from playing $(DE,DS)$, we should have
 
\begin{align}
\label{ineq1}
V_{\omega_{(DE,DS)}}^{D} > (1-\delta)u_{D}(DE,S) + \delta V_{\omega_{(E,DS)}}^{D} 
\end{align}
 
where the right-hand-side of the inequality is the utility from deviating to playing $ S $ for one period and then transitioning to the state $V_{\omega_{E,DS}}$. Here, the value for the inital state is simply $V_{\omega_{DE,DS}}^{D} = -50$, i.e. the drivers' stage game utility for $(DE, DS)$, since following the automaton the action profile $(DE,DS)$ is repeated infinitely implying (\ref{geom_sum}) becomes $(1-\delta) \cdot (-50) \cdot 1/(1-\delta) = -50$. To calculate $V_{\omega_{(E,DS)}}$, we use the payoff function and follow the automaton:
 
\begin{align}
V_{\omega_{(E,DS)}}^{D} &= (1-\delta)u_{D}(E,DS) + \delta V_{\omega_{(DE,DS)}}^{D} \\
                 &= - 50 \cdot (1-\delta) + (-50)\delta \\
                 &= -50.
\end{align}

If inequality (\ref{ineq1}) held, the drivers would never have an incentive to speed since being taken on the punishment path - here simply being punished with $(E,DS)$ for one round (year) - would be too costly. However,  
 
\begin{align}
V_{\omega_{(NE,DS)}}^{D} &> (1-\delta)u_{D}(DE,S) + \delta V_{\omega_{(E,DS)}}^{D}\\
  \Leftrightarrow -50 &> (1-\delta ) \cdot 50 + \delta \cdot (-50)\\
  \Leftrightarrow -50 &> 50 - 100\delta \\ \Leftrightarrow \delta &> 1, 
\end{align}
 
so this is shown to be impossible due to $\delta \in [ 0,1 ] $. We reach the same (fairly obvious) conclusion as before: given the stage-game utilities given as in Elvik (2015)\cite{elvik}, it's not possible to take away the drivers' incentive to speed with the threat of punishing them with traffic enforcement the next year for the duration of one year only.
 
A more reasonable attempt to construct a two-state automaton describing an SPE is to postulate the initial state as a mixed strategy state where the police are enforcing with a certain probability (or frequency) and the drivers speeding with another probability. As noted earlier, there exists a mixed strategy equilibrium for the stage game where the drives are speeding 50 $\%$ of the time. Then repeating this equilibrium infinitely gives us a very simple example of the kind of subgame equilibria we're looking for. Remember, however, that this equilibrium is viewed as unstable by Bjørnskau $\&$ Elvik (1992)\cite{elvik0}. This implies that any stable equilibrium outcome we postulate should at least have the property that the drivers are speeding clearly less than 50 $\%$ of the time.

Let us denote the mixed-strategy-state (which is also the initial state) by $\omega_{ms}$ and the probability of the drivers speeding by $b_D(S)$ as before. The punishment state remains $\omega_{(E,DS)}$. We next ask the following question: if the police was not enforcing \emph{at all} (i.e. $b_P(E) = 0$), meaning speeding becomes as attractive as possible and the intial state as valuable as possible for the drivers, how large could $b_D(S)$ be in an SPE? For the remainder of this article, we will make the assumption of $b_P(E) = 0$ which, if supportable, would be the optimal equilibrium outcome for the police. 
 
For there to be no incentive for the drivers to deviate from the state $V^{D}_{\omega_{ms}}$, the following has to hold:

\begin{align}
V^{D}_{\omega_{ms}} &> (1-\delta ) u_{D}(DE,S) + \delta V^{D}_{\omega_{(E,DS)}}\\
							&= (1-\delta )u_{D} (DE,S) + \delta ((1-\delta )u_{D}(E,DS) + \delta V_{\omega_{ms}}^{D}) \label{autopluspayoff}\\
\Leftrightarrow (1-\delta^2)V_{\omega_{ms}}^{D} &> 50(1-\delta ) -50\delta (1-\delta ) \\
V_{\omega_{ms}}^{D} &> 50 \frac{1-\delta }{1+\delta }
\end{align}

where we have used the payoff function twice to get (\ref{autopluspayoff}). Since $ V_{\omega_{ms}}^{D} = 50b_D(S) + (-50)(1-b_D(S)) = 100b_D(S) - 50$ when the police is playing pure strategy $DE$ and the drivers are mixing, the criterion for deviation becomes
 
\begin{align}
100b_D(S)-50 &> 50\frac{1-\delta}{1+\delta} \\ \Leftrightarrow b_D(S) &> \frac{1}{1+\delta} \label{ongelma} 
\end{align}
 
This implies $1/2 < b_D(S)< 1$, so we'd have to accept quite a high value for $b_D(S)$ to achieve an SPE strategy profile. However, the police can quarantee a payoff of at least $-10000$ by just playing $E$ every period, so at the very least we'd need to have
 
\begin{align}
\centering
V_{\omega_{ms}}^{P} &> -10^{4} \\
\Leftrightarrow -2\cdot 10^{4}b_{D}(S) + 0\cdot (1-b_{D}(S)) &> -10^{4} \\ \Leftrightarrow b_{D}(S) &< \frac{1}{2}.
\end{align}  
 
\begin{figure}
\begin{center}
\begin{tikzpicture}[scale=1.1, every node/.style={scale=1.1}]
\node[state, initial] (q2) {$\hspace{1.0em}\omega_{\textrm{ms}\hspace{1.0em}}$};
\node[state, right of=q2, xshift=3cm] (q3) {$\hspace{0.5em}\omega_{E,DS}\hspace{0.5em}$};
\draw[->] (q2) edge[loop above] node{\scriptsize $(DS, (b_D(S),1-b_D(S)))$} (q2);
\draw[->] (q2) edge[bend left, above] node{\scriptsize $b_D(S) \geq 1/2$}(q3);
\draw[->] (q3) edge[bend left, below] node{\scriptsize (E,DS)}(q2);
\draw[->] (q3) edge[loop above] node{\scriptsize not (E,DS)} (q3);
\end{tikzpicture}
\end{center}
\caption{\small{Automaton ii. $\omega_{ms}$ is the initial state. This profile also fails to be an SPE.}}
\label{auto2}
\end{figure}
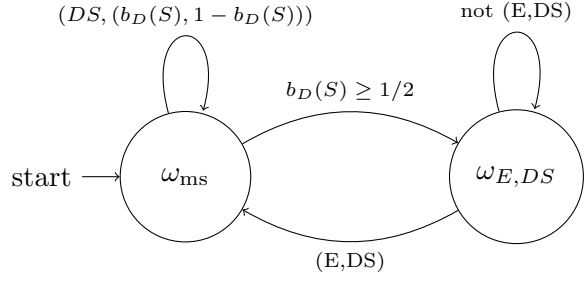

But this contracicts (\ref{ongelma}) and shows it's not possible to have the automaton in Figure \ref{auto2} depicting an SPE in this case.

This is a meaningful finding in itself. We have discovered that in addition to the drivers having an incentive to speed, the police has \emph{no incentive} to take away the drivers' incentive to speed by threatning to punishment them with enforcement for one year. One-shot deviations always exist for both parties given the stage-game matrix as in Elvik (2015)\cite{elvik}.
 
To try to construct an SPE strategy profile, we could investigate the possibility of having a more elaborate punishment path - the drivers being punished for more than one period - but if a period corresponds to one year, it wouldn't seem feasible to punish the drivers for multiple years. Given that the police is able to change their strategy quicker (as discussed earlier), we are going to suggest the possibility of shortening the period duration. This would imply assessing the data multiple times a year. Traffic safety work of the police has both strategic and operational dimensions (Poliisi, 2022: 24)\cite{poliisi2022}. The police can use monthly reports on traffic enforcement and/or traffic offences and crashes in their traffic analysis at operational (or tactical) level but one year can be more relevant for strategic level. Thus, the police are able to change their strategy somewhat easily to some extent. Now, if there are multiple "checkpoints" a year, the stage-game utilities should also vary. In addition to decreasing the magnitude off the stage-game utilities, we - more importantly  - also suggest generalizing their proportions. For this, suppose that in the new stage-game the negative utilites for the police are replaced with functions of $b_D(S)$. After all, the more speeding takes place, the greater should be the extent of enforcement and the greater the potential damage done by speeding. We therefore replace $-10000$ with $-\beta(b_D(S))$ and $-20000$ with $\alpha(b_D(S))$ ($\alpha$ and $\beta$ taking only positive values so that the minus sign exits to remind us the utilites for the police are non-negative). When the duration of one period is shortened to one $N$th of of a year ($N\in \mathds{N}\setminus \{1\}$), the drivers' utilites are divided by $N$. Also, the discount factor should be $N$-dependent such that it it's an increasing function of $\delta_N$ of $N$.

\begin{table}
\label{huihaiii}
\centering
   \setlength{\extrarowheight}{2pt}
   \begin{tabular}{cc|c|c|}
     & \multicolumn{1}{c}{} & \multicolumn{2}{c}{Drivers}\\
     & \multicolumn{1}{c}{} & \multicolumn{1}{c}{$S$}  & \multicolumn{1}{c}{$DS$} \\\cline{3-4}
     \multirow{2}*{Police}  & $E$ & $-\beta(b_D(S)),-300/N$ & $-\beta(b_D(S)),-50/N$ \\\cline{3-4}
     & $NE$ & $-\alpha(b_D(S)),50/N$ & $0,-50/N$ \\\cline{3-4}
   \end{tabular}
\caption{\small{Stage-game matrix for short-period repeated game}} 
\end{table}
 
Let $n<N$ be the number of periods the driver is going to be punished for deviating from $\omega_{ms}$. The punishment stage-game action profile remains as before, $(E,DS)$. The police then values being in the state $\omega_{ms}$ as
 
$$V^{P}_{\omega_{ms}} = -\alpha (b_D(S))b_D(S) + 0 \cdot (1-b_D(S)) = -\alpha (b_D(S))b_D(S)$$
 
since the police is playing a pure strategy and the drivers are mixing. For the drivers, this value is
 
$$
V^{D}_{\omega_{ms}} = 50 \frac{1}{N} \cdot b_D(S) + (-50 \frac{1}{N}) \cdot (1-b_D(S)) = \frac{1}{N} (100b_D(S)-50).
$$
 
Since the new punishment path has $n$ states, $i$th state denoted $\omega_{(E,DS)_{i}}$, the value for the first state on that path for the drivers is

\begin{align}
V_{\omega_{(E,DS)_1}}^D &= (1-\delta _N)((-50\frac{1}{N}) + (-50\frac{1}{N})\delta _N + (-50\frac{1}{N}) \delta _N ^2 +\\ &\hspace{1em}\cdots + (-50\frac{1}{N})\delta_N ^{n-1})+ \delta _N ^n V_{\omega_{ms}}^D\\
&= -(50\frac{1}{N})(1-\delta _N ^n) + \delta _N ^n (100\frac{1}{N} b_D(S) - 50\frac{1}{N} ) \\
&= \frac{1}{N}(100\delta _N ^n b_D(S) -50)
\end{align}

When it comes to the drivers, we in fact only need this value with respect to the states on the punishment path to check for one-shot deviations. This is because the drivers will not have an incentive to deviate from playing $DS$ whenever the police plays $E$. As for the requirement for the drivers to have no incentive to deviate from $\omega_{ms}$, the following must hold:
 
\begin{align}
\frac{1}{N}(100 b_D(S) - 50) &> (1-\delta _N)u_D(DE,S) + \delta _N V_{(E,DS)_1}^D \\ 
&= \frac{1}{N} (50(1-\delta _N ) + \delta _N (100 \delta _N ^n b_D(S)-50)) \\ 
&= \frac{1}{N} (100 \delta _N  ^{n+1} b_D(S)-100 \delta _N + 50) \\ 
\Leftrightarrow (100-100\delta _N ^{n+1})b_D(S) &> -100 \delta _N + 100 \\ 
\Leftrightarrow b_D(S) &> \frac{1-\delta _N }{1-\delta  _N ^{n+1}}. 
\label{ehtodeltan} 
\end{align}

\setcounter{footnote}{0}

So we observe that $b_D(S)$ is dependent on both the discount factor and the number of periods. We can study\footnote{An illustrative way would be to do this graphically, but we leave it for the reader.} this dependence by varying $n$. Clearly we'd like $n$ to be as small as possible to maximize the feasibility of the punishment. If $n=1$, (\ref{ehtodeltan}) becomes $\frac{1}{2} < b_D(S) < 1$, so in this case the drivers would have to be speeding at least half of the time, which is clearly suboptimal. If $n=2$ and $\delta$ is close to $1$, it's possible to have $b_D(S)$ nearly as small as $0,33$ (this is because $\lim_{\delta _N \rightarrow 1} (1-\delta _N)/(1-\delta _N ^ 3) = \lim_{\delta _N\rightarrow 1} 1/(\delta _N ^2 + \delta _N +1) = \frac{1}{3}$). If $n=3$, the lower limit becomes as low as $0,20$. The implication here is that the police could allow the probability of speeding only somewhere near this lower limit and use the punishment path whenever the drivers speed at a higher probability.

How should $N$ be chosen in the light of these values? If we fix $N = 365$, the punishment lasts for $n$ days. If $N = 12$, the punishment lasts for $n$ months. With respect to our model of perfect information, we could make $N$ enourmosly large with the following consequences: $n$ becomes miniscule resulting in a punishment path of extremely short duration (e.g. one nanosecond), the drivers become extremely patient ($\delta _N \approx 1$) and the drivers still have no incentive to deviate from $\omega _{ms}$. Then again, if the drivers had an incentive to deviate even for a nanosecond, they would do this continuously, and nanoseconds would eventually turn into minutes, hours, days, weeks and months and so on. We therefore make the observation that, when it comes to the drivers' incentives, the duration of a period is not relevant (within the model). However, from the perspective of the police there are at least two externalities that become limiting factors here, namely the time it takes for the police to react to the drivers' deviation and the resources the police requires in order to carry out the punisment for the deviation.
 
To investigate the incentives of the police, we first calculate how the police values being in $\omega_{(E,DS)_1}$:

\begin{align}
V_{\omega_{(E,DS)_{1}}}^{P} &= (1-\delta _N)(-\beta (b_D(S)) + (-\beta (b_D(S)))\delta _N + (-\beta (b_D(S))) \delta _N ^2 \label{ehtopolincentive_0}\\
+&\cdots + (-\beta (b_D(S))) \delta _N ^{n-1}) + \delta _N ^{n} V^{P}_{ms} \\
&= -\beta (b_D(S))(1-\delta _N ^n) + \delta _N ^n V^P_{\omega_{ms}}\\
&= -\beta (b_D(S))(1-\delta _N ^n) + \delta _N ^n (-\alpha (b_D(S))b_D(S)), \label{ehtopolincentive_1}
\end{align}
 
This means the police has no incentive to deviate from the initial state $\omega_{ms}$ given that
 
\begin{align}
-\alpha (b_D(S)) b_D(S) &> (1-\delta _N) (-\beta (b_D(S))) + \delta _N V_{\omega_{(E,DS)_1}} ^ P \label{polekapunish_0}\\ 
&= (1-\delta _N) (-\beta (b_D (S))\\ &\hspace{2em}+\delta _N (-\beta (b_D(S))(1-\delta _N ^ n) \\ 
&\hspace{2em} +\delta _N ^n (-\alpha (b_D(S))b_D(S))) \\ 
\Leftrightarrow -\alpha (b_D(S))b_D(S) (1-\delta _N ^{n+1}) &>  (\delta _N (1-\delta ^n) + (1-\delta _N ))(-\beta (b_D(S))) \\ 
\Leftrightarrow -\alpha (b_D(S)) b_D(S) &> \frac{(\delta _N (1-\delta _N ^n) + (1-\delta _N))(-\beta (b_D(S)))}{(1-\delta _N ^{n+1})}\\ 
\Leftrightarrow b_D(S) &< \frac{(\delta _N (1-\delta _N ^n) + (1-\delta _N))}{1-\delta _N ^{n+1}} \frac{\beta (b_D(S))}{\alpha ( b_D(S)) } \\ 
&= \frac{\beta (b_D(S))}{\alpha (b_D(S))} 
\end{align}
 
As this condition holds if and only if the probability of speeding $b_D(S)$ must be lower than the ratio $\beta / \alpha $ at $b_D(S)$, there is an incentice to bring the cost of enforcement $\beta$ as low as possible with respect to the cost of damages from not enforcing $\alpha$, quite intuitively.

\begin{figure}
\begin{center}
\begin{tikzpicture}[scale=1, every node/.style={scale=1}]
\node[state, initial](q1){$\hspace{1.0em}\omega_{\textrm{ms}}\hspace{1.0em}$};
\node[state, right of=q1, xshift =3.0cm] (q4) {$\hspace{0.2em}\omega_{(\textrm{\tiny E,DS})_1}\hspace{0.2em}$};
\node[state, right of=q4, xshift=1.7cm] (q2) {$\hspace{0.2em}\omega_{(\textrm{\tiny E,DS})_2}\hspace{0.2em}$};
\node[state, right of=q2, xshift=2.5cm] (q3) {$\hspace{0.2em}\omega_{(\textrm{\tiny E,DS})_n}\hspace{0.2em}$};
\draw[->] (q1) edge[loop above] node{\textrm{\scriptsize $(DS, (b_D(S), 1-b_D(S)))$}} (q1);
\draw[->] (q1) edge[above] node{\textrm{\scriptsize $b_D(S) > \frac{1-\delta _N}{1-\delta _N ^n} + \epsilon $}} (q4);
\draw[->] (q2) edge[loop above] node{\textrm{\scriptsize not (E,DS)}} (q2);
\draw[->] (q4) edge[above] node{\textrm{\scriptsize (E,DS)}} (q2);
\draw[->] (q2) edge[above, middle dotted line={line width=0.4pt}] node{} (q3);
\draw[->] (q3) edge[bend left, below] node{\textrm{\scriptsize (E,DS)}} (q1);
\draw[->] (q3) edge[loop above] node{\textrm{\scriptsize not (E,DS)}} (q3);
\draw[->] (q4) edge[loop above] node{\textrm{\scriptsize not (E,DS)}} (q4);
\end{tikzpicture}
\caption{\footnotesize{Automaton iii. $n$-state punishment path. This profile is a NE but not an SPE.} }
\end{center}
\end{figure}
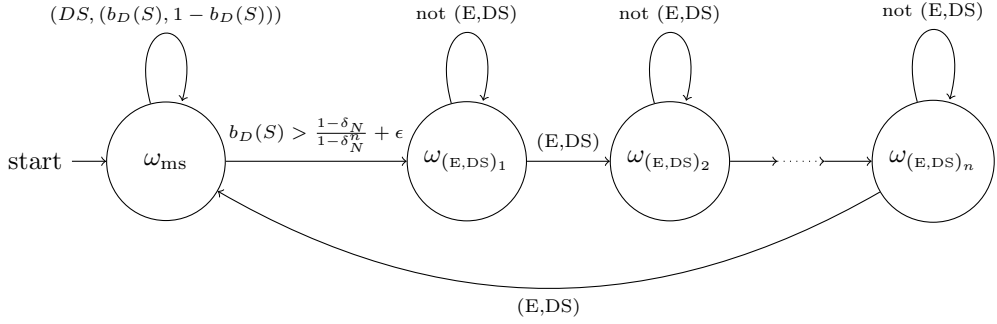
 
Aside from externalities, there exists a bigger problem within the model when it comes to the incentives of the police. Turns out that even though the police has no incentive to deviate from $\omega_{ms}$ (all the other stage game pure strategy outcomes yield a lower utility for the police), it has an incentive to deviate from \emph{any} state on the punishment path. Take the first state on the punishment path for example. The police values being in $\omega_{(E,DS)_1}$ as $ - \beta (1 - \delta ^n) + \delta _N ^n (-\alpha (b_D(S))b_D(S)) $ (see (\ref{ehtopolincentive_0})-(\ref{ehtopolincentive_1})) which is always non-positive given the definitions of $\alpha$ and $\beta$. In order for the police to not have an incentive to deviate from playing $(E,DS)$ when in $\omega_{(E,DS)_1}$ would mean that the police prefers $(DE,DS)$ to be played and being taken back to $\omega_{(E,DS)_1}$ by the transfer rule to following the automaton. We thus should have

\begin{align}
V_{\omega_{(E,DS)_1}}^P &> (1-\delta _N) u_P(DE,DS) + \delta _N V_{\omega_ {(E,DS)_1}}^P\\
&= (1-\delta _N) \cdot 0 + \delta _N V_{\omega_ {(E,DS)_1}}^P \\
\Leftrightarrow (1-\delta _N) V_{\omega_{(E,DS)_1}}^P &> 0\\
\Leftrightarrow V_{\omega _{(E,DS)_1}}^P &> 0,
\end{align}

which is not possible given (\ref{ehtopolincentive_0})-(\ref{ehtopolincentive_1}). We have shown that the police has a profitable deviation from a state of the automaton and therefore it cannot depict an SPE (we could have used any state on the punishment path instead of $\omega_{(E,DS)_1}$ to similarly show this).
 
Because neither the police or the drivers have an incentive to deviate from the inital state but there are states on the punishment path that one of the players has an incentive to deviate from, the automaton actually depicts a Nash equilibrium but not a subgame perfect equilibrium. The implication is that if the drivers act irrationally (assuming they are more prone to doing so) and deviate from the initial state by speeding with too high a probability (frequency), the police would need to be funded to have an incentive to carry out the punishment or they would have to be legally obligated to carry out the punishment regardless of the expenses. Certain externalities, perhaps especially the concept of deterrence as discussed earlier, may come to the aid of the police here. Even if the drivers had an incentive to deviate from ''not speeding'' on the punishment path in a very technical sense, the may be psychologically discouraged to do so. Better yet, deterrence together with the game theoretic disincentive should, in general, contribute to keeping the drivers in the inital state $\omega_{ms}$ in the first place.
 
Out of interest, we examine the possibility of there existing a third party that gives financial aid to the police whenever the drivers behave irrationally and deviate from $V_{\omega_{ms}}$. We denote the smallest sum that is needed to fund the police with to prevent them from having an incentive to deviate from any state on the punishment path with $\gamma $. Notice here that the police is funded with $\gamma$ \emph{only if} the drivers deviate from the intial state and there is no other way for the police to get this funding. It should also be forbidden to use $\gamma$ for anything other than covering the expenses for traffic enforcement. Then the payoff for the police from $(E,S)$ becomes $-\beta(b_D(S)) + \gamma /n$ - that is, one $n$th of $\gamma$ is added to the stage game utility.
 
Given this funding, the police has no incentive to deviate from playing $E$ given that
 
\begin{equation}
(1-\delta _N) (-\beta _D (S) + \gamma / n) (1 + \delta _N + \delta _N ^2 + \cdots + \delta _N ^{n-m}) + \delta ^{n-m+1} V_{\omega_{ms}} ^P \geq 0 
\end{equation}

for all $m \in {1,...,n}$. After rearranging the terms, we can easily see that this condition is equivalent to
 
\begin{align}
\gamma / n &\geq \sup _{m\leq n} \biggl\{ \Bigl(1-\frac{\delta _N  ^{n-m+1}}{1-\delta _N ^{n-m+1}}\Bigr)\beta (b_D(S)) \biggl\} \\ 
&= \Bigl(1-\frac{\delta _N ^{n}}{1-\delta _N ^{n}}\Bigr)\beta (b_D(S)), 
\end{align}
 
and thus at the optimum
 
\begin{align}
\gamma = \frac{1}{n} \Bigl(1-\frac{\delta _N ^{n}}{1-\delta _N ^{n}}\Bigr)\beta (b_D(S)).
\end{align}

\hspace{1em}

\begin{center}
\textmd{\small CONCLUSIONS}
\end{center}

Even though subgame perfect strategy profiles always exist in infinitely repeated games, constructing an SPE profile that regulates the drivers' incentive to speed in the most desirable way is not a trivial task. We saw that it is impossible to have a stable solution with a strategy profile having only one state on its punishment path given that the probability of speeding is at least 50 percent while the police is controlling. A strategy profile of an infinitely repeated game with an initial state where the police is also mixing - i.e. not enforcing some of the time and enforcing rest of the time - was not discussed in this paper (except the repetition of the stage game mixed strategy NE), which leaves room for further analysis and possibly discovery of game theoretically stable solution that is also societally feasible. An outcome in a ``short period repeated game'' where the drivers are speeding with a probability $b_D(S)$ bounded by the ratio $\beta(b_D(S)) / \alpha(b_D(S))$, the functions $\beta$ and $\alpha$ generalizing the stage game payoffs, was shown to be supportable as a Nash equilibrium but not an SPE. This has the implication that when a player enters a certain subgame (or subgames), perhaps willingly - yet irrationally - or by a mistake, the player no longer has an incentive to follow the strategy profile, which means the player's behaviour drifts farther from the social planner's control. Certain externalities, such as deterrence, would be likely to help to maintain a stable outcome in a real-life setting. 

It is clear that if the drivers are to be punished in some way or sense for violating the traffic laws, the punishment policy and its extent can't be just arbitrarly guessed and implemented. Rather, it should stem from a carefully premeditated analysis that follows rationally solid argumentation which, of course, should further be evaluated and refined in the light of empirical evidence. In this article we have demonstrated how game theoretic modeling and the solution concept of SPE is one way to find basis for such logical arguments.

Our infinitely repeated game model was restricted to the case of a driver having perfect information. Even though it is very unrealistic that a driver would calculate his or her expected utility whenever entering a vehicle in a very complex fashion as we have done in this article, the repeated game framework is able to show that \emph{even if} the driver did this with perfect information, she or he could not do any better in an SPE than by following the equilibrium strategy profile. Any deviation from the driver makes the situation worse for the driver, which is a very appealing feature of the Nash equilibrium concept. For further research, one might want to consider introducing elements of uncertainty to the repeated game. Another good question is how to properly define the utility generalizing functions $\alpha$ and $\beta$. They could be made state-dependent such that the state is charaterized by weather, region, time of the year, traffic jams, traffic culture etc. The same can be said about the discount factor $\delta$ that needn't be common among the players. The model could also be made dependent of available traffic data. For example, the behaviour strategy of the driver(s) could follow a period-wise stochastic process that fits real-life data. This could help in predicting the next point of time when the drivers are about to irrationally exit the inital mixed strategy state in our short period repeated game.

\end{document}